\documentclass[10pt]{iopart}
\usepackage{latexsym}
\usepackage{amsfonts}
\usepackage{amsbsy}
\usepackage{amssymb}   
\usepackage{mathrsfs}
\usepackage[mathscr]{euscript}
\usepackage{accents}
\usepackage{hyperref}
\hypersetup{colorlinks=true,linkcolor=blue,urlcolor=blue,citecolor=blue}

\usepackage[makeroom]{cancel}

\usepackage{tensor}
\usepackage{color}
\usepackage{cite}

\newcommand{\eqref}[1]{(\ref{#1})}
\newcommand{\overbar}[1]{\mkern 1.5mu\overline{\mkern-1.5mu#1\mkern-1.5mu}\mkern 1.5mu}

\newcommand{\intprod}{\mbox{$\; \put(0,0){\line(1,0){.9}}
		\put(.9,0){\line(0,1){1.6}} \; \, \,  $}}

\begin{document}

\title[Trace-free Einstein gravity as two interacting constrained $BF$ theories]
{Trace-free Einstein gravity as two interacting constrained $BF$ theories}

\author{Merced Montesinos$^{1}$  and Diego Gonzalez$^{1,2}$}

\address{$^1$ Departamento de F\'{\i}sica, Cinvestav, Avenida Instituto Polit\'ecnico Nacional 2508, 	San Pedro Zacatenco, 07360 Gustavo A. Madero, Ciudad de M\'exico, M\'exico.}
\address{$^2$ Escuela Superior de Ingenier\'ia Mec\'anica y El\'ectrica, Instituto Polit\'ecnico Nacional, Unidad Profesional Adolfo L\'opez Mateos, Zacatenco,  07738 Gustavo A. Madero, Ciudad de M\'exico, M\'exico}

\eads{\mailto{merced.montesinos@cinvestav.mx} and \mailto{dgonzalezv@ipn.mx}}


\begin{abstract}

A theory of gravity alternative to general relativity is trace-free Einstein gravity, which has the remarkable property that the cosmological constant emerges as an integration constant. In this paper, we report two fully diffeomorphism-invariant actions for trace-free Einstein gravity. They describe the theory as two $BF$ theories supplemented with some constraints. The first action comprises two copies of the constrained $BF$ theory for the Husain-Kucha\v{r} model plus an interaction term involving the fields that impose the constraints on the $B$ fields. The second action employs two copies of the chiral Plebanski action for general relativity plus an additional constraint. Both actions use complex variables, and naturally include one of the reality conditions imposed in the Plebanski formulation of general relativity. The new actions have the advantage of not involving any nondynamical fields or unimodular condition, and their only gravitational sector is trace-free Einstein gravity.

\end{abstract}


\section{Introduction}

Trace-free Einstein gravity is an alternative theory of gravity that replaces the equations of general relativity by their trace-free part~\cite{Einstein_1919,Einstein_1952,Einstein_1927}, with the advantage that it solves the fundamental problem of vacuum energy~\cite{Ellis_2011,Ellis_2014}. The idea is that the trace-free Einstein equations, along with the second Bianchi identity and the assumption of the conservation of the energy-momentum, bring back the equation of general relativity involving the trace, but with the cosmological constant emerging as an integration constant unrelated to vacuum energy. Although conceptually very deep, fully diffeomorphism-invariant actions for trace-free Einstein gravity have only recently come to light by the authors in~\cite{MontGonz_2023}. These actions describe trace-free Einstein gravity as a real $BF$ theory\footnote{The term `$BF$ theory' originates from the specific form of the action principle, which involves the wedge product of the 2-form $B$ and the curvature $F$.} with a constraint on the $B$ field, without involving any nondynamical fields or imposing the unimodular condition present in the so-called unimodular gravity~\cite{Anderson_1971,Weinberg_1989}, where the general diffeomorphism invariance is broken~\cite{Carballo-Rubio_2022,Bengochea_2023}. Remarkably, one of these actions includes a free parameter that does not modify the equations of motion of trace-free Einstein gravity. Furthermore, the actions given in~\cite{MontGonz_2023} encompass not only trace-free Einstein gravity but also a gravitational sector corresponding to Einstein’s general relativity with vanishing cosmological constant. While not a drawback, a fully diffeomorphism-invariant formulation having  only trace-free Einstein gravity as the gravitational sector would be desirable.

The aim of this paper is to further investigate formulations of trace-free Einstein gravity and to present two fully diffeomorphism-invariant actions that employ complex variables, and in which the gravitational sector of Einstein's general relativity is absent. The new actions reported here describe trace-free Einstein gravity for Lorentzian signature as two $BF$ theories supplemented with constraints on the fields. We use the same reality conditions that are usually employed in the Plebanski formulation of general relativity~\cite{pleb1977118} (see also~\cite{BFgravity}). However, the way they enter is different from one formulation to the other.

One of the features that makes our actions worth considering is their very simple form. The first action can be regarded as two copies of the constrained $BF$ theory for the Husain-Kucha\v{r} model introduced in~\cite{MontVel_2009}, coupled with an interaction term involving the fields that impose the constraints on the $B$ fields. Remarkably, it is the interaction term that prevents the theory from being degenerate, as happens in the Husain-Kucha\v{r} model alone~\cite{HusainKucharmodel}. Another remarkable feature of this formulation is that it inherently involves one of the reality conditions. The second action comprises two copies of the chiral Plebanski action for general relativity, plus an interaction term that explicitly involves one of the reality conditions. It is noteworthy that this interaction term plays a double role, introducing a reality condition, and preventing the theory from describing general relativity, as is the case with the standalone Plebanski action. Furthermore, being of the $BF$-type, our actions could be suitable for exploring the quantum aspects of trace-free Einstein gravity within the framework of spin foam models~\cite{RovBook,ThieBook,Perez_2013,livine2024spinfoam}.

The paper is organized as follows. In section~\ref{section_TFEG}, we provide a brief review of trace-free Einstein gravity in the first-order formalism~\cite{MontGonz_2023}. In section~\ref{section_action}, we present the first action and show that its equations of motion are equivalent to the trace-free Einstein equations for Lorentzian signature. We also show how, in our formulation, the cosmological constant emerges as an integration constant due to the Bianchi identity. In section~\ref{Equiv_action}, we present the second action. We also show its equivalence with trace-free Einstein gravity and how the cosmological constant arises. Finally, in section~\ref{section_conclusions}, we present our conclusions and directions for future research.

\section{Tetrad and connection formalism of trace-free Einstein gravity}\label{section_TFEG}

The aim of this section is to recall some relevant aspects of trace-free Einstein gravity in the first-order formalism. Let us begin by fixing the notation. In this formalism, the metric is $g =\eta_{IJ} e^I \otimes e^J$ where $(\eta_{IJ}) := \mbox{diag} (\sigma,1,1,1)$ with $\sigma=-1$ for the Lorentzian and $\sigma=1$ for the Euclidean metrics; $e^I$ is the dual basis of the orthonormal frame $\partial_I$ in the sense that $e^I (\partial_J)= \delta^I_J$. Thus, $g (\partial_I, \partial_J) = \eta_{IJ}$. The other fundamental variable is the connection $\omega^I{}_J$, which is compatible with the metric and torsion-free 
\begin{eqnarray}
D\eta_{IJ}:=d\eta_{IJ}-\omega^K{}_I\eta_{KJ}-\omega^K{}_J \eta_{IK}=0, \label{spinc1} \\
D e^I := d e^I + \omega^I{}_J \wedge e^J=T^I, \label{spinc2}
\end{eqnarray}
where $d$ is the exterior derivative and $\wedge$ is the wedge product and $T^I$ is the torsion 2-form. Thus, the capital indices $I, J, K, \ldots=0,1,2,3$ are raised (lowered) with $\eta^{IJ}$ ($\eta_{IJ}$), and additionally $\varepsilon_{IJKL}$ is the totally anti-symmetric $SO(3,1)$ invariant tensor if $\sigma=-1$ [$SO(4)$ if $\sigma=1$] with $\varepsilon_{0123}=1$. 

In the absence of matter fields, and assuming vanishing torsion,
\begin{equation}\label{notorsion}
D e^I =0,
\end{equation}
the trace-free Einstein equations can be written in the first-order formalism as~\cite{MontGonz_2023}
\begin{equation}\label{eq_forms}
\ast \mathcal{R}_{IJ}=\star \mathcal{R}_{IJ},
\end{equation}
where $\ast$ is the internal dual and $\star$ is the Hodge dual of the curvature of $\omega^I{}_J$, given by $\mathcal{R}^I{}_J = d \omega^I{}_J + \omega^I{}_K \wedge \omega^K{}_J$. The internal dual and the Hodge dual of ${\mathcal R}_{IJ}$ are a priori two different objects. If they are equal to each other, then the equations for trace-free Einstein gravity emerge. That is the beauty of the formulation given by~\eqref{eq_forms}. In fact, the internal dual is simply $\ast {\mathcal R}_{IJ}= \frac12 \varepsilon_{IJ}{}^{KL} {\mathcal R}_{KL}$. The Hodge dual, on the other hand, is a map that sends $p$-forms to $(4-p)$-forms. Thus, writing the curvature ${\mathcal R}_{IJ}$ in the basis $e^I$ as ${\mathcal R}_{IJ} = \frac12 R_{IJKL} e^K \wedge e^L$, we have $\star {\mathcal R}_{IJ} = \frac12 R_{IJKL} \star \left (e^K \wedge e^L \right)= \frac14 R_{IJKL} \varepsilon^{KL}{}_{MN} e^M \wedge e^N = \frac12 {R \ast}_{IJKL} e^K \wedge e^L$. Therefore, equation~\eqref{eq_forms} amounts to
\begin{equation}\label{1_eq_forms}
{\ast R}_{IJKL} = {R \ast}_{IJKL},
\end{equation}
which is equivalent to
\begin{equation}\label{2_eq_forms}
{\ast R \ast}_{IJKL} = \sigma R_{IJKL}.
\end{equation}
However, ${\ast R \ast}_{IJKL}$ is given by
\begin{eqnarray}\label{3_eq_forms}
\fl {\ast R \ast}_{IJKL} =& \sigma R_{KLIJ} 
+ \sigma \bigg[ \left( R_{LI} -\frac14 R \eta_{LI} \right)\eta_{JK} - \left( R_{KI} -\frac14 R \eta_{KI} \right)\eta_{JL} \nonumber\\
\fl &+\left(R_{KJ} -\frac14 R \eta_{KJ} \right)\eta_{LI} -\left(R_{LJ} -\frac14 R \eta_{LJ} \right)\eta_{KI} \bigg],
\end{eqnarray}
where $R_{IJ}=R^K{}_{IKJ}$ are the components of the Ricci tensor and $R=R^{IJ}{}_{IJ}$ is the scalar curvature. Substituting \eqref{3_eq_forms} in~\eqref{2_eq_forms} and using the fact that $R_{IJKL}=R_{KLIJ}$---which comes from the first Bianchi identity $\mathcal{R}^I{}_J \wedge e^J =0$---and taking an appropriate trace, we get the trace-free Einstein equations in the familiar form  
\begin{equation}\label{TFEM2}
R_{IJ} -\frac 14 R \eta_{IJ}=0.
\end{equation}
The converse, namely, that \eqref{TFEM2} implies \eqref{eq_forms} can also be shown along the same lines.

On the other hand, using~\eqref{spinc1},~\eqref{notorsion},~\eqref{eq_forms}, and the second Bianchi identity $D {\mathcal R}^I{}_J=0$, it is deduced that
\begin{equation}
d R =0,  \label{dR}
 \end{equation}
which entails
\begin{equation}\label{origin_Lambda}
  R =: 4 \Lambda,  
\end{equation}
where $\Lambda$ is an integration constant. Therefore, as has been pointed out previously, in trace-free Einstein gravity the cosmological constant $\Lambda$  arises as an integration constant implied by the second Bianchi identity.

An alternative way to express the trace-free Einstein equations is as follows. The starting point is the observation that the thirty six components of the curvature $R^{IJ}{}_{KL}$ give rise to four $3\times 3$ matrices: $E=(E^i{}_j)$, $H=(H^i{}_j)$, $Q=(Q^i{}_j)$, and $F=(F^i{}_j)$ with components $E^i{}_j :=R^{0i}{}_{0j}$, $H^i{}_j :=\frac12 \varepsilon_j{}^{kl} R^{0i}{}_{kl}$, $Q^i{}_j :=\frac{\sigma}{2} \varepsilon^i{}_{kl} R^{kl}{}_{0j}$, and $F^i{}_j :=\frac14 \varepsilon^i{}_{kl} \varepsilon_j{}^{mn}  R^{kl}{}_{mn}$, where $i,j,k,\ldots=1,2,3$ and $\varepsilon_{ijk}$ is the totally anti-symmetric invariant $SO(3)$ tensor ($\varepsilon_{123}=1$) with $\varepsilon_{0ijk}=\varepsilon_{ijk}$ and $\varepsilon^{0ijk}=\sigma\varepsilon^{ijk}$. In terms of these matrices, the trace-free Einstein equations~\eqref{eq_forms} are equivalent to $F=E$ and $Q=H$. However, because of the first Bianchi identity, these matrices have the properties $E= E^T$, $F= F^T$, $H= Q^T$, and ${\rm Tr}H= 0$, and hence another way to write the trace-free Einstein equations is
\begin{equation} \label{mtx_TFG_1}
F=E, \quad E=E^T, \quad Q= H, \quad H= H^T, \quad {\rm Tr}H= 0. 
\end{equation}
Therefore, the trace-free Einstein equations imply that the Riemann tensor acquires the following form 
\begin{equation}
\left ( R^{IJ}{}_{KL} \right ) = \left( \begin{array}{cc}
E & H \\
\sigma H  & E  \end{array} \right),
\end{equation}
with 
\begin{equation}
E= E^T, \quad H = H^T, \quad {\rm Tr}\, H=0.
\end{equation}
In this context, $R=4 {\rm Tr}\,E$ and thus
\begin{equation}\label{origin_Lambda_2}
{\rm Tr}\, E = \Lambda,
\end{equation}
on account of~\eqref{origin_Lambda}.

\section{The action}\label{section_action}
In this section we show that in the absence of matter fields, trace-free Einstein gravity can be described as two copies of the constrained $BF$ theory that yields the Husain-Kucha\v{r} model reported in~\cite{MontVel_2009} plus an interaction term involving the fields that impose the constraints on the $B$ fields. More precisely, the first-order action for trace-free Einstein gravity is given by
\begin{eqnarray}\label{complexBF_TFG}
\fl S\left[A,\overbar{A},\Sigma,\overbar{\Sigma},\Psi,\overbar{\Psi},\rho\right] =&  \int \Big[ \Sigma_i \wedge F^{i} -\frac12 \Psi_{ij} \Sigma^{i} \wedge \Sigma^{j} + \overbar{\Sigma}_i \wedge \overbar{F}^{i} 
-\frac12 \overbar{\Psi}_{ij} \overbar{\Sigma}^i \wedge \overbar{\Sigma}^j \nonumber \\
&-\rho \left( {\rm Tr}\Psi - {\rm Tr}\overbar{\Psi} \right) \Big],
\end{eqnarray}
where $\Sigma^i$  is a triad of $\mathfrak{so}(3,\mathbb{C})$-valued 2-forms, $F^i=dA^i+\frac12 \varepsilon^i{}_{jk} A^j \wedge  A^k$ is the curvature of the  SO$(3,\mathbb{C})$-connection $A^i$, $\Psi=(\Psi^i{}_j)$ is a symmetric complex $3\times 3$  matrix, and $\rho$ is a nonvanishing 4-form that plays the role of a Lagrange multiplier imposing the constraint that ${\rm Tr}\Psi$ equals ${\rm Tr}\overbar{\Psi}$. This condition is just a single real constraint. Therefore, it is reasonable to take $\rho=- \overbar{\rho}$ to keep the action real. Here, an overbar denotes complex conjugate, and we use the internal indices $i,j,k,\ldots=1,2,3$, which are raised and lowered with the three dimensional Euclidean metric~$\delta_{ij}$.

The equations of motion that follow from~\eqref{complexBF_TFG} are
\begin{eqnarray}
    \delta \Psi_{ij} \,&:\,\, \Sigma^i \wedge \Sigma^j + 2 \rho \delta^{ij}=0, \label{EOM_plus_psi} \\
    \delta A^i \,&:\,\, D_A \Sigma^i := d \Sigma^i + \varepsilon^i{}_{jk} A^j \wedge \Sigma^k=0, \label{EOM_plus_A}\\
    \delta \Sigma^i \,&:\,\, F^i = \Psi^i{}_j \Sigma^j, \label{EOM_plus_sigma} \\
    \delta \overbar{\Psi}_{ij} \,&:\,\, \overbar{\Sigma}^i \wedge \overbar{\Sigma}^j - 2 \rho \delta^{ij}=0, \label{EOM_neg_psi} \\
    \delta \overbar{A}^i \,&:\,\, D_{\overbar{A}} \overbar{\Sigma}^i := d \overbar{\Sigma}^i + \varepsilon^i{}_{jk} \overbar{A}^j \wedge \overbar{\Sigma}^k=0, \label{EOM_neg_A} \\
    \delta \overbar{\Sigma}^i \,&:\,\, \overbar{F}^i = \overbar{\Psi}^i{}_j \overbar{\Sigma}^j, \label{EOM_neg_sigma} \\
    \delta \rho \,&:\,\, {\rm Tr}\Psi - {\rm Tr}\overbar{\Psi} = 0, \label{EOM_rho}
\end{eqnarray}
where $D_A$ and $D_{\overbar{A}}$ are the covariant derivatives defined by $A^i$ and $\overbar{A}^i$, respectively. In what follows, we will show that these equations are equivalent to the trace-free Einstein equations for Lorentzian signature ($\sigma=-1$).

We begin from \eqref{EOM_plus_psi}. Computing the trace of these equations to get rid of $\rho$, we get the solution for $\Sigma^i$~\cite{pleb1977118,Capovilla_1991,BFgravity}
\begin{equation}
    \Sigma^i = \tau \left( {\rm i} \kappa e^0 \wedge e^i - \frac12 \varepsilon^{i}{}_{jk} e^j \wedge e^k \right) = \frac{\tau}{2} P^i{}_{JK} e^J \wedge e^K, \label{sol_plus_sigma}
\end{equation}
where $\tau=\pm1$, $P^i{}_{JK}:={\rm i} \kappa ( \delta^0_J \delta^i_K - \delta^i_J \delta^0_K) + \varepsilon^{0i}{}_{JK}$ is the self-dual projector for $\kappa=1$ (respectively, anti-self-dual for $\kappa=-1$), and $\{e^I\}=\{e^0,e^i\}$ is a set of four linearly-independent complex 1-forms. Substituting this solution in~\eqref{EOM_plus_psi}, the 4-form $\rho$ acquires the following form
\begin{equation}
    \rho=-\frac16 \Sigma_i \wedge \Sigma^i = {\rm i} \kappa e^0 \wedge e^1 \wedge e^2 \wedge e^3. \label{sol_plus_rho}
\end{equation}

To obtain trace-free Einstein gravity for Lorentzian signature, it is necessary to consider {\it reality conditions}. Such conditions are the same as those imposed in the Plebanski formulation to recover general relativity for Lorentzian signature~\cite{pleb1977118} (see also~\cite{BFgravity}), namely 
\begin{eqnarray}
    \Sigma_i \wedge \Sigma^i + \overbar{\Sigma}_i \wedge \overbar{\Sigma}^i=0, \label{real_cond1} \\
    \Sigma^i \wedge \overbar{\Sigma}^j =0. \label{real_cond2}
\end{eqnarray}
Note that \eqref{real_cond1} follows from computing the traces of \eqref{EOM_plus_psi} and \eqref{EOM_neg_psi}, after getting rid of $\rho$.  It is noteworthy that our formulation inherently includes this reality condition. On the other hand, the conditions \eqref{real_cond2} must be additionally imposed, just as in the case of the Plebanski formulation (or any other complex formulation) of general relativity. With the conditions \eqref{real_cond1} and \eqref{real_cond2} at hand, the 1-forms $e^I$ involved in \eqref{sol_plus_sigma} can be taken to be real, and we will take them as real from now on.

 On account of~\eqref{sol_plus_sigma}, the solution of~\eqref{EOM_plus_A} is
\begin{equation}
    A^i = \frac12 P^i{}_{JK} \omega^{JK}, \label{sol_plus_A}
\end{equation}
with the connection $\omega^I{}_J$ satisfying~\eqref{spinc1} and~\eqref{notorsion}. In turn, the expression \eqref{sol_plus_A} implies that 
\begin{equation}
    F^i = \frac12 P^i{}_{JK} \mathcal{R}^{JK}, \label{sol_plus_F}
\end{equation}
where $\mathcal{R}^I{}_J$ is the curvature of $\omega^I{}_J$. 

Then, using $\mathcal{R}^{IJ}=\frac12 R^{IJ}{}_{KL} e^K \wedge e^L $ and the identity $e^K \wedge e^L= \frac{\tau}{2} ( P_i{}^{KL} \Sigma^i + \overbar{P}_i{}^{KL} \overbar{\Sigma}^i)$, we rewrite $\mathcal{R}^{IJ}$ as 
\begin{equation}
\mathcal{R}^{IJ} = \frac{\tau}{4} R^{IJ}{}_{KL} \left ( P_i{}^{KL} \Sigma^i + \overbar{P}_i{}^{KL} \overbar{\Sigma}^i \right ). \label{identidad}
\end{equation}
Using this expression, equation~\eqref{sol_plus_F} becomes
\begin{equation}
F^i = \frac{\tau}{8} P^i{}_{IJ} R^{IJ}{}_{KL} \left ( P_j{}^{KL} \Sigma^j  + \overbar{P}_j{}^{KL} \overbar{\Sigma}^j \right ). \label{sol_plus_F_2}
\end{equation}
Therefore, substituting this result in the left-hand side of~\eqref{EOM_plus_sigma}, we arrive at
\begin{eqnarray}
    \frac{\tau}{8} P^i{}_{JK} P_j{}^{MN} R^{JK}{}_{MN} = \Psi^{i}{}_j, \label{proy_plus_1} \\
    P^i{}_{JK} \overbar{P}_j{}^{MN} R^{JK}{}_{MN} = 0. \label{proy_plus_2}
\end{eqnarray}
To better understand the meaning of these equations, it is convenient to express them in  terms of the matrices $E$, $H$, $Q$, and $F$. By doing so, equations~\eqref{proy_plus_1} and \eqref{proy_plus_2}  read
\begin{eqnarray}
     \frac{\tau}{2} \left[F+E - {\rm i} \kappa (Q+H) \right]  = \Psi, \label{mtx_plus_1}\\
    F-E + {\rm i} \kappa (Q-H)  = 0, \label{mtx_plus_2}
\end{eqnarray}
respectively. We thus see that \eqref{mtx_plus_2} implies 
\begin{equation}
  F=E, \quad Q=H. \label{mtx_TFG}   
\end{equation}
Furthermore, using \eqref{mtx_TFG}, equation~\eqref{mtx_plus_1} leads to
\begin{equation}
    E=\frac{\tau}{2} (\Psi + \overbar{\Psi}), \quad H=\frac{ {\rm i} \tau \kappa}{2} (\Psi - \overbar{\Psi}), \label{mtx_TFG2}  
\end{equation}
from which it is clear that $E=E^T$ and $H=H^T$. Additionally, this means that \eqref{EOM_rho} amounts to 
\begin{equation}
 {\rm Tr}H= 0. \label{Binchi_trH}   
\end{equation}
Remarkable, equations~\eqref{mtx_TFG} together with $E=E^T$, $H=H^T$, and~\eqref{Binchi_trH} are precisely the trace-free Einstein equations~\eqref{mtx_TFG_1}. In this way, we have shown that the action~\eqref{complexBF_TFG} describes trace-free Einstein gravity for Lorentzian signature.

The same conclusion can be drawn by following an analogous path using the complex conjugate of the variables. We can readily see that the complex conjugate of \eqref{sol_plus_sigma} and \eqref{sol_plus_A} are respectively the solutions of \eqref{EOM_neg_psi} and \eqref{EOM_neg_A}. Then, as expected, the curvature of $\overbar{A}^i$ is given by the complex conjugate of \eqref{sol_plus_F_2}, which together with \eqref{EOM_neg_sigma} leads to the complex conjugate of \eqref{mtx_plus_1} and \eqref{mtx_plus_2}, giving rise to \eqref{mtx_TFG} and \eqref{mtx_TFG2}, as before.  

{\it Meaning of $\Psi$ and $\overbar{\Psi}$}. From \eqref{proy_plus_1}, it is direct to see that, for $\kappa=1$, $\Psi$ and $\overbar{\Psi}$ are related to the self-dual and anti-self-dual parts of the Riemann tensor $R^{IJ}{}_{KL}$, respectively. For $\kappa=-1$, $\Psi$ and $\overbar{\Psi}$ are related to the anti-self-dual and self-dual parts of the Riemann tensor $R^{IJ}{}_{KL}$, respectively. Therefore, in this formulation of trace-free Einstein gravity, the meaning of $\Psi$ and  $\overbar{\Psi}$ is different from that of the analogous matrices involved in the Plebanski formulation of general relativity, which are related to the self-dual and anti-self-dual parts of the Weyl tensor~\cite{pleb1977118} (see also~\cite{BFgravity}). 


{\it Cosmological constant as an integration constant.} Taking the covariant derivative $D_A$ of \eqref{EOM_plus_sigma} and using \eqref{EOM_plus_A} as well as the Bianchi identity $D_A F^i=0$, we get
\begin{equation}
    \left ( D_A \Psi^i{}_j \right ) \wedge \Sigma^j=0. \label{cosmo_3F}
\end{equation}
Contracting this expression with the coordinate basis $\partial_\alpha$ of the tangent space of spacetime and then making the wedge product of the resulting expression with $\Sigma_i$, we obtain
\begin{equation}
    [ \partial_\alpha \intprod \left ( D_A \Psi^i{}_j \right ) ]\Sigma^j  \wedge \Sigma_i - D_A \Psi^i{}_j \wedge ( \partial_\alpha \intprod \Sigma^j) \wedge \Sigma_i =0. \label{cosmo_4F}
\end{equation}
We recall that the symbol $\intprod$ stands for the contraction of a vector with a differential form and that $\partial_\alpha$ is the dual basis of $dx^\alpha$ (i.e., $\partial_\alpha \intprod dx^\beta = \delta^\beta_\alpha$). Now, using \eqref{EOM_plus_psi} and \eqref{cosmo_3F}, equation~\eqref{cosmo_4F} reduces to $\partial_\alpha \intprod \left ( d {\rm Tr}\Psi \right )=0$, which implies
\begin{equation}
    d{\rm Tr}\Psi=0. \label{cosmo_dTrplus}
\end{equation}

On the other hand, by taking the trace of \eqref{proy_plus_1} and using the first Bianchi identity, we get 
\begin{equation}\label{result}
{\rm Tr}\Psi=\frac{\tau}{4} R.
\end{equation}
Consequently, \eqref{cosmo_dTrplus} and~\eqref{result} imply 
\begin{equation} \label{differential_R}
    d R= 0,
\end{equation}
from which the cosmological constant $\Lambda$ emerges as an integration constant (see equations~\eqref{origin_Lambda} and~\eqref{origin_Lambda_2}). 

It is worth noting that due to \eqref{EOM_rho}, equation~\eqref{cosmo_dTrplus} amounts to $d{\rm Tr}\overbar{\Psi}=0$, which also implies \eqref{differential_R} since the complex conjugate of~\eqref{result} yields ${\rm Tr}\overbar{\Psi}=\frac{\tau}{4} R$.

\section{Equivalent action}\label{Equiv_action}

We now present an action for trace-free Einstein gravity that is equivalent to \eqref{complexBF_TFG}. The action consists of two copies of the chiral Plebanski action, coupled with an interaction term involving the reality condition~\eqref{real_cond1}. It is given by
\begin{eqnarray}\label{complexBF_TFG_2}
\fl S\left[A,\overbar{A},\Sigma,\overbar{\Sigma},\Phi,\overbar{\Phi},\nu,\overbar{\nu},\mu\right] 
=&  \int \Big[ \Sigma_i \wedge F^{i}  -\frac12 \Phi_{ij} \Sigma^{i} \wedge \Sigma^{j} -\nu  {\rm Tr}\Phi 
+ \overbar{\Sigma}_i \wedge \overbar{F}^{i} \nonumber \\
&-\frac12 \overbar{\Phi}_{ij} \overbar{\Sigma}^i \wedge \overbar{\Sigma}^j  - \overbar{\nu} {\rm Tr}\overbar{\Phi} 
-\frac{\mu}{2} \left( \Sigma_{i} \wedge \Sigma^{i} + \overbar{\Sigma}_i \wedge \overbar{\Sigma}^i \right)\Big],
\end{eqnarray}
where $\Phi=(\Phi^i{}_j)$ is a symmetric complex $3\times 3$  matrix, $\nu$ is a nonvanishing 4-form Lagrange multiplier, and  $\mu$ is a Lagrange multiplier imposing \eqref{real_cond1}. As before, an overbar denotes complex conjugate, and then we take $\mu=\overbar{\mu}$ to keep the action real. It is not hard to show that performing the redefinition $\Phi^i{}_j=\Psi^i{}_j - \mu \delta^i{}_j$, the action~\eqref{complexBF_TFG_2} leads to \eqref{complexBF_TFG} with $\nu=\rho$.

Alternatively, the equations of motion arising from~\eqref{complexBF_TFG_2} are
\begin{eqnarray}
    \delta \Phi_{ij} \,&:\,\, \Sigma^i \wedge \Sigma^j + 2 \nu \delta^{ij}=0, \label{EOM_plus_Phi_2} \\
    \delta A^i \,&:\,\, D_A \Sigma^i =0, \label{EOM_plus_A_2}\\
    \delta \Sigma^i \,&:\,\, F^i = \Phi^i{}_j \Sigma^j + \mu \Sigma^i, \label{EOM_plus_sigma_2} \\
    \delta \nu \,&:\,\, {\rm Tr}\Phi = 0, \label{EOM_plus_nu_2} \\
    \delta \overbar{\Phi}_{ij} \,&:\,\, \overbar{\Sigma}^i \wedge \overbar{\Sigma}^j + 2 \overbar{\nu} \delta^{ij}=0, \label{EOM_neg_Phi_2} 
    \\
    \delta \overbar{A}^i \,&:\,\, D_{\overbar{A}} \overbar{\Sigma}^i=0, \label{EOM_neg_A_2} \\
    \delta \overbar{\Sigma}^i \,&:\,\, \overbar{F}^i = \overbar{\Phi}^i{}_j \overbar{\Sigma}^j + \mu \overbar{\Sigma}^i, \label{EOM_neg_sigma_2} \\
    \delta \overbar{\nu} \,&:\,\, {\rm Tr}\overbar{\Phi} = 0, \label{EOM_plus_nu_2} \\
    \delta \mu \,&:\,\, \Sigma_{i} \wedge \Sigma^{i} + \overbar{\Sigma}_i \wedge \overbar{\Sigma}^i = 0. \label{EOM_plus_nu_2} 
\end{eqnarray}
To show that these equations are equivalent to the trace-free Einstein equations for Lorentzian signature ($\sigma=-1$), we can follow a procedure similar to the one described in the previous section. After getting rid of $\nu$, equation~\eqref{EOM_plus_Phi_2} together with \eqref{EOM_plus_nu_2} and the conditions \eqref{real_cond2} imply that the 2-forms $\Sigma^i$ are given by \eqref{sol_plus_sigma} with real 1-forms $e^I$. Because of this the solution of \eqref{EOM_plus_A_2} is the connection~\eqref{sol_plus_A}, whose curvature can be expressed as \eqref{sol_plus_F_2}. Then, substituting~\eqref{sol_plus_F_2} in \eqref{EOM_plus_sigma_2}, we get 
\begin{eqnarray}
    \frac{\tau}{8} P^i{}_{JK} P_j{}^{MN} R^{JK}{}_{MN} = \Phi^{i}{}_j + \mu \delta^i_j, \label{Equv_proy_plus_1} \\
    P^i{}_{JK} \overbar{P}_j{}^{MN} R^{JK}{}_{MN} = 0. \label{Equv_proy_plus_2}
\end{eqnarray}
Notice that \eqref{Equv_proy_plus_2} is exactly the same as \eqref{proy_plus_2}, thereby implying \eqref{mtx_TFG}. Along with $E=E^T$, $H=H^T$, and ${\rm Tr}H= 0$ (which come from the first Bianchi identity because $\omega^I{}_J$ has no torsion), these are the trace-free Einstein equations~\eqref{mtx_TFG_1}. The same conclusion can be reached by taking a similar path with the complex conjugates variables.

Also, notice that the meaning of $\Phi$ and $\overbar{\Phi}$ is different from that of the matrices $\Psi$ and $\overbar{\Psi}$ involved in the formulation~\eqref{complexBF_TFG}. In fact, using \eqref{Equv_proy_plus_1}, it turns out that, for $\kappa=1$, $\Phi$ and $\overbar{\Phi}$ related to the self-dual and anti-self-dual parts of the Weyl tensor, respectively. For $\kappa=-1$, the roles of $\Psi$ and $\overbar{\Psi}$ are exchanged. 

We conclude this section by showing how the cosmological constant emerges as an integration constant in this formulation. Using \eqref{EOM_plus_A_2} and $D_A F^i=0$, the covariant derivative $D_A$ of \eqref{EOM_plus_sigma_2} leads to
\begin{equation}
    \left ( D_A \Phi^i{}_j \right ) \wedge \Sigma^j + d\mu\wedge\Sigma^i=0. \label{cosmo_3F_2}
\end{equation}
Contracting \eqref{cosmo_3F_2} with the coordinate basis of the tangent space of spacetime $\partial_\alpha$ and then making the wedge product with $\Sigma_i$, we get
\begin{eqnarray}
    [ \partial_\alpha \intprod \left ( D_A \Phi^i{}_j \right ) ]\Sigma^j  \wedge \Sigma_i - D_A \Phi^i{}_j \wedge ( \partial_\alpha \intprod \Sigma^j) \wedge \Sigma_i \nonumber \\
    +(\partial_\alpha \intprod d\mu) \Sigma^i \wedge \Sigma_i - d\mu \wedge ( \partial_\alpha \intprod \Sigma^i) \wedge \Sigma_i = 0. \label{cosmo_4F_2}
\end{eqnarray}
After using \eqref{EOM_plus_Phi_2}, \eqref{EOM_plus_nu_2}, and \eqref{cosmo_3F_2}, this expression reduces to $\partial_\alpha \intprod d\mu=0$, which entails
\begin{equation}
    d\mu=0. \label{dmu}
\end{equation}
In addition, the trace of \eqref{Equv_proy_plus_1} together with the first Bianchi identity (we recall $\omega^I{}_J$ has no torsion) yields 
\begin{equation}
    R=12 \tau \mu.
\end{equation}
From this and \eqref{dmu} it is clear that $dR=0$, giving rise to the cosmological constant $\Lambda$ (see equations~\eqref{origin_Lambda} and~\eqref{origin_Lambda_2}). 

\section{Conclusions}\label{section_conclusions}

In this paper, we have presented two new fully diffeomorphism-invariant actions for trace-free Einstein gravity for Lorentzian signature. They describe trace-free Einstein gravity as two $BF$ theories supplemented with some constraints: the first action uses two copies of the constrained $BF$ theory for the Husain-Kucha\v{r} model reported in~\cite{MontVel_2009} plus an interaction term involving the fields that impose the constraints on the $B$ fields, while the second action employs two copies of the chiral Plebanski action for general relativity plus an additional constraint. In this way, each formulation offers a different perspective on trace-free gravity, revealing unanticipated connections with other theories. Both actions involve complex variables, and so we used reality conditions on the $B$ fields. These conditions are the same as those imposed in the Plebanski formulation of general relativity. Notably, one of these conditions is included in both formulations; however, it enters in different ways.  The first action incorporates this condition implicitly, while the second does it explicitly through the interaction term. This fact can be considered as a step forward compared to the Plebanski formulation of general relativity, where the complete set of reality conditions is imposed by hand. Clearly, the price we pay for this is that our actions are not chiral. A remarkable aspect of the current formulations is that they do not involve the general relativity sector, in contrast with the real formulation reported in~\cite{MontGonz_2023} (see, however,~\cite{Mont_Gonz_bigravity} for a new formulation of trace-free Einstein gravity including matter
fields as a constrained bigravity theory that also does not involve the general relativity sector).

It is important to emphasize that our actions do not employ nondynamical fields or the unimodular condition to yield the trace-free Einstein equations, thereby preserving the full diffeomorphism invariance of the theory. Furthermore, in our formulations the cosmological constant emerges as a consequence of the Bianchi identity, in line with Einstein's original proposal~\cite{Einstein_1919,Einstein_1927}. This is in contrast with the actions introduced in~\cite{Gielen_2024}, which do not describe diffeomorphism-invariant trace-free Einstein gravity, but rather unimodular gravity. 

Finally, our formulations for trace-free Einstein gravity may pave new avenues for exploring both classical and quantum aspects of gravity. On the classical side, given that the canonical analysis of the Plebanski action leads directly to the Ashtekar formalism of general relativity~\cite{Capovilla_1991}, it would be very valuable to study the Hamiltonian formulation of these actions along the lines of~\cite{MontVel_2009,BFgravity,Montesinos_HBF_2020,Montesinos_HBF_2021} (and references therein). This would also be of great interest for loop quantum gravity~\cite{RovBook,ThieBook,Ashtekar_2021}. Moreover, on the quantum side, they serve as a promising starting point for exploring the spinfoam program for quantum gravity \cite{RovBook,ThieBook,Perez_2013,livine2024spinfoam}, offering a tentative path integral quantization of trace-free Einstein gravity, which could be confronted with quantum unimodular gravity~\cite{Smolin_2009,Smolin_2011}.

\ack 

Diego Gonzalez acknowledges the financial support of Instituto Polit\'ecnico Nacional, Grant No. SIP-20240184, and the postdoctoral fellowship from Consejo Nacional de Humanidades, Ciencia y Tecnología (CONAHCyT), México.


\section*{References}
\bibliography{references}

\end{document}